\documentclass[reprint,noshowpacs,noshowkeys,prd,balancelastpage,nofootinbib]{revtex4}
\usepackage{amsfonts}
\usepackage{mdframed}
\usepackage{amssymb}
\usepackage{footnote}
\usepackage{amsmath}
\usepackage{graphicx}
\usepackage{float}
\usepackage[font={footnotesize,it}]{caption}
\usepackage[utf8]{inputenc}
\usepackage{natbib}
\usepackage{tikz}
\usepackage{tikz-3dplot}
\usepackage[colorlinks=true,
            linkcolor=purple,
            urlcolor=purple,
            citecolor=blue]{hyperref}

\begin{document}

\title{Quark-antiquark confinement and nonlinear electrodynamics}
\author{S. Habib Mazharimousavi}
\email{habib.mazhari@emu.edu.tr}
\affiliation{Department of Physics, Faculty of Arts and Sciences, Eastern Mediterranean
University, Famagusta, North Cyprus via Mersin 10, Turkiye}
\date{\today }

\begin{abstract}
Cornell potential is known to represent the quark-antiquark confinement
interaction. In addition to the Cornell potential, there have been other
interactions in the literature that demonstrate confining structure in the
quark-antiquark system. Guendelman has proposed a nonlinear electrodynamics
(NED) model which results in electric potential in the form of Cornell
interaction. In this study, we propose two NED models whose electric
potential is in the form of two other confining potentials. We also study
the coupling of gravity to one of these models which are in the form of a
correction to linear Maxwell's theory. We find spacetime which can be a
black hole or a naked singular particle. In the zeroth approximation, we
examine the circular speed of a distant test particle orbiting the central
black hole. The Newtonian potential due to the spacetime at the location of
the distant particle possesses an additional term proportional to the NED
parameter. Due to the presence of such a term, the speed-radius curve gets
improved from the pure Reissner-Nordstrom. It is a kind of weaker
confinement acting on distant parts of a galaxy due to a supermassive
central black hole.
\end{abstract}

\maketitle

\section{Introduction}

In quantum chromodynamics the mass spectrum and electromagnetic transitions
of the charmonium/quark-antiquark have been theoretically determined in the
seminal work of Eichten, et al in Ref. \cite{1} where the so-called Cornell
potential was proposed (see also \cite{C1,C2,C3,C4,C5}). In accordance with 
\cite{1}, the spherically symmetric Cornell potential consists of two parts,
a color Coulomb term and a confining linear term expressed by%
\begin{equation}
V_{c}\left( r\right) =-\frac{\alpha _{s}}{r}+\frac{\alpha _{s}}{a^{2}}r
\label{1}
\end{equation}%
in which $\alpha _{s}$ is the charmonium fine structure constant ($\simeq
0.2 $) and $a$ is a dimensionful parameter ($\simeq 0.2$ fm). The structure
of the Cornell potential is such that at the small distance between the
quark-antiquark the Coulomb term dominants i.e., $V\left( r\right)
\rightarrow -\frac{\alpha _{s}}{r}$ as $r\rightarrow 0$ and for a large
distance, the linear term dominants i.e., $V\left( r\right) \rightarrow 
\frac{\alpha _{s}}{a^{2}}r$ as $r\rightarrow \infty $.

In addition to the Cornell potential, there have been other proposed
potentials for such systems in the literature which have shown improvement
in the agreement of the theory and observation. One of a rather early
potential after \cite{1} was proposed in \cite{2} where the potential
between quark and antiquark was expressed by%
\begin{equation}
V_{q}\left( r\right) =C\ln \left( \frac{r}{a}\right)  \label{2}
\end{equation}%
where $C$ and $a$ are dimensionful parameters. Both potentials confine the
quark-antiquark system such that the system remains definite-bounded.
Moreover, considering the vacuum polarization effect of the dynamical light
quark suggests \cite{S1,S2,S3,S4,S5} a fine-tuned correction to the Cornell
potential in the form of a screened potential such that the potential
overall may look like 
\begin{equation}
V\left( r\right) =-\frac{\alpha _{s}}{r}+\frac{b}{\mu }\left( 1-\exp \left(
-\mu r\right) \right)  \label{SC}
\end{equation}%
in which $b$ and $\mu $ are constant parameters and $\mu $ is the screening
factor. The potential (\ref{SC}) behaves Coulombic at small $r,$ however, at
very large $r$ such that $\mu r\gg 1$ becomes a constant while in the near
zone with $\mu r\ll 1$ it reduces to the Cornell potential. 
\begin{figure}[tbp]
\includegraphics[scale=0.7]{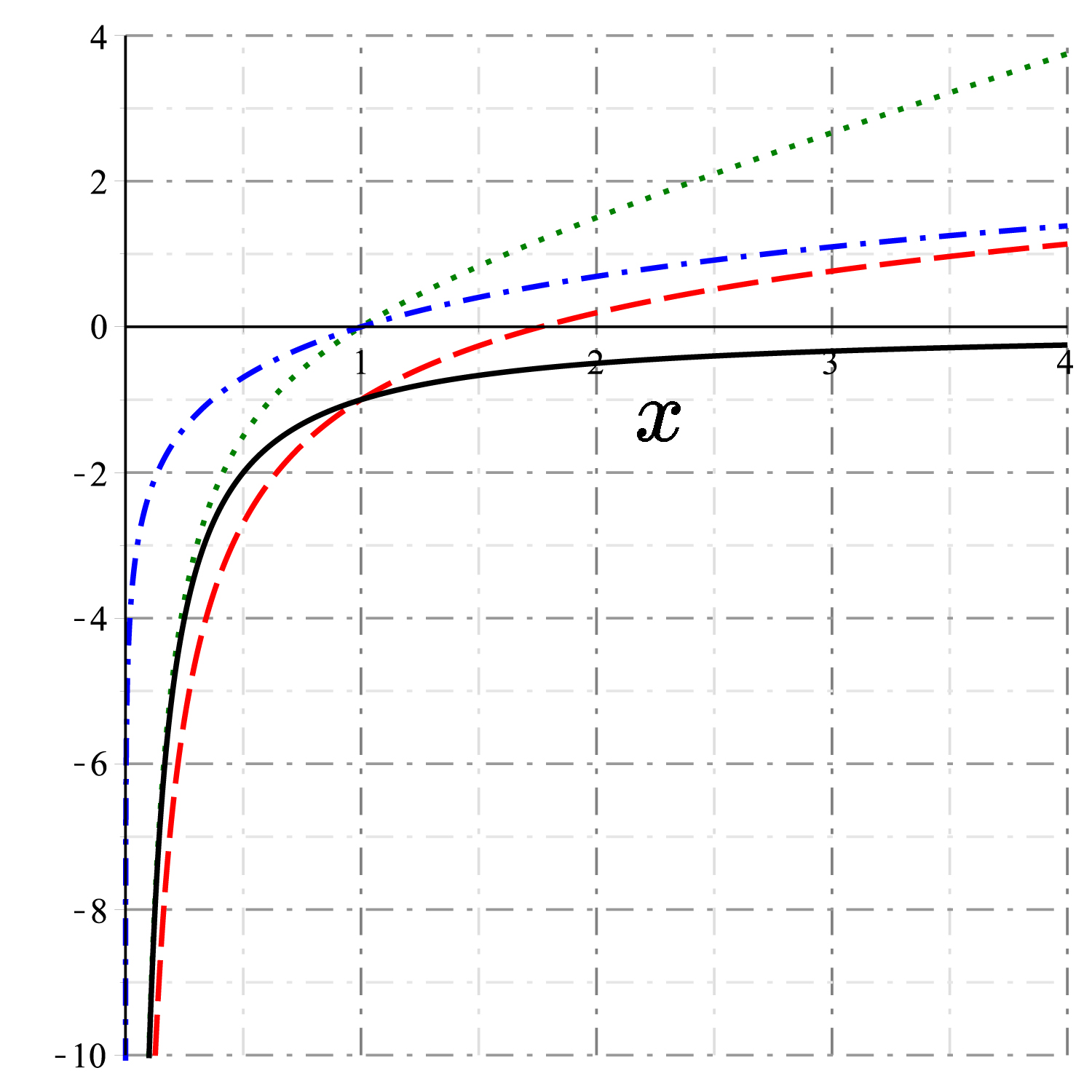}
\caption{Plots of $\frac{a}{\protect\alpha _{s}}V_{c}$ (green dot curve),$%
\frac{1}{C}V_{q}$ (blue dash-dot curve), and $-\frac{a}{r}$ (black solid
curve) versus $x=\frac{r}{a}$. Furthermore, the red-long dash curve
represents $\frac{a}{\protect\alpha }V_{h}$ with respect to $x$. }
\label{Fig1}
\end{figure}

In Fig. \ref{Fig1}, we plotted $\frac{a}{\alpha _{s}}V_{c}$ (green dot
curve),$\frac{1}{C}V_{q}$ (blue dash-dot curve), and $-\frac{a}{r}$ (black
solid curve) versus $x=\frac{r}{a}$. It is observed that while $V_{c}$
agrees with the Coulomb potential at a small $r$, $V_{q}$ doesn't. In Fig. %
\ref{Fig1} we also plotted $\frac{a}{\alpha }V_{h}$ where 
\begin{equation}
V_{h}\left( r\right) =-\frac{\alpha }{r}+\frac{\alpha }{a}\ln \left( \frac{r%
}{a}\right) ,  \label{3}
\end{equation}%
which displays significant improvement and all three potentials i.e., $%
V_{c},V_{q},$ and $V_{h}$ agree at a small $r$. Here in this study, we are
not going to solve the Schr\"{o}dinger equation for the quarkonium with
potential (\ref{3}) and obtain the bound spectrum. Instead, we are going to
look at the interaction of the quark and antiquark in the framework of
nonlinear electrodynamics (NED) such that the corresponding nonrelativistic
potential is expressed as either (\ref{3}) or (\ref{2}).

\section{Nonlinear electrodynamics: Cornell potential}

Regarding to the charmonium potential (\ref{1}) Guendelman in Ref. \cite{3}
(see also \cite{4}) assumed a NED nature for the Cornell potential. In \cite%
{3} and \cite{4} (see also \cite{R1}) the following NEDs model has been
proposed 
\begin{equation}
\mathcal{L}=-\mathcal{F}-f\sqrt{-\mathcal{F}}  \label{4}
\end{equation}%
where $\mathcal{F}=\frac{1}{4}F_{\mu \nu }F^{\mu \nu }$ is the Maxwell
invariant, $\mathcal{L}$ is the NED Lagrangian and $f$ is a coupling
constant. For a point charge sitting at the origin, in the differential form
notation, we write%
\begin{equation}
\mathbf{F=}E\left( r\right) dt\wedge dr  \label{5}
\end{equation}%
in which $E\left( r\right) $ is the radial electric field in the flat
spacetime described by the following line element ($c=1$)%
\begin{equation}
ds^{2}=-dt^{2}+dr^{2}+r^{2}\left( d\theta ^{2}+\sin ^{2}\theta d\varphi
^{2}\right) .  \label{6}
\end{equation}%
The field tensor (\ref{5}) satisfies the Bianchi identity i.e., $d\mathbf{F}%
=0$, however, Maxwell's nonlinear equation implies%
\begin{equation}
d\left( \mathbf{\tilde{F}}\frac{\partial \mathcal{L}}{\partial \mathcal{F}}%
\right) =0  \label{7}
\end{equation}%
in which 
\begin{equation}
\mathbf{\tilde{F}=}E\left( r\right) r^{2}\sin \theta d\theta \wedge d\varphi
\label{8}
\end{equation}%
is the Hodge-dual of $\mathbf{F}$. Maxwell's equation explicitly yields%
\begin{equation}
E\left( r\right) r^{2}\frac{\partial \mathcal{L}}{\partial \mathcal{F}}=C
\label{9}
\end{equation}%
where $C$ is an integration constant. Knowing that $\mathcal{F}=-\frac{1}{2}%
E^{2}$ one finds from (\ref{9}) 
\begin{equation}
E\left( r\right) =\frac{q}{r^{2}}+\frac{f}{\sqrt{2}}  \label{10}
\end{equation}%
where we set $C=-q$ to get the correct Maxwell's linear theory in the limit $%
f\rightarrow 0$. The electric potential corresponding to the radial electric
field (\ref{10}) is found to be%
\begin{equation}
U\left( r\right) =\frac{q}{r}-\frac{f}{\sqrt{2}}r.  \label{11}
\end{equation}%
This electric potential in the NED model (\ref{4}) clearly gives the same
charmonium interaction potential energy upon a proper adjustment of
parameters. In an analogy with Guendelman's approach, in the sequel, we
investigate the NED model corresponding to the other two interaction
potentials introduced in (\ref{2}) and (\ref{3}).

\section{Nonlinear electrodynamics: Coulomb with Logarithmic correction
interaction}

In this section, we start with the confinement potential (\ref{3}) and
introduce an electric potential in the form%
\begin{equation}
U\left( r\right) =\frac{q}{r}-f\ln \left( \frac{r}{r_{0}}\right)  \label{12}
\end{equation}%
where we assume both $q$ and $f$ are positive. This yields a radial electric
field given by%
\begin{equation}
E\left( r\right) =\frac{q}{r^{2}}+\frac{f}{r}.  \label{13}
\end{equation}%
Next, we consider Maxwell's equation with an unknown NED Lagrangian (\ref{9}%
) and obtain the following differential equation to be satisfied by $%
\mathcal{L}$%
\begin{equation}
\frac{d\mathcal{L}}{dr}=-\frac{q}{r^{4}}\left( \frac{2q}{r}+f\right)
\label{14}
\end{equation}%
where for the same reason as (\ref{10}), we set $C=-q.$ The latter equation
yields a solution for $\mathcal{L}\left( r\right) $ given by%
\begin{equation}
\mathcal{L}=\frac{q^{2}}{2r^{4}}+\frac{qf}{3r^{3}}+\mathcal{L}_{0}
\label{15}
\end{equation}%
in which $\mathcal{L}_{0}$ is an integration constant. Next we use (\ref{13}%
) to get%
\begin{equation}
\mathcal{F}=-\frac{1}{2}\left( \frac{q}{r^{2}}+\frac{f}{r}\right) ^{2}
\label{16}
\end{equation}%
upon which we eliminate $r$ from $\mathcal{L}$ in (\ref{15}) to get%
\begin{equation}
\mathcal{L}=-\frac{16\left( 3\sqrt{-2\mathcal{F}}+\zeta \left( \zeta +\sqrt{%
\zeta ^{2}+4\sqrt{-2\mathcal{F}}}\right) \right) \sqrt{-2\mathcal{F}}}{%
3\left( \zeta +\sqrt{\zeta ^{2}+4\sqrt{-2\mathcal{F}}}\right) ^{4}}\mathcal{F%
}  \label{17}
\end{equation}%
where we have set $\mathcal{L}_{0}=0$ and $f=\zeta \sqrt{q}$ with $\zeta $ a
new constant parameter. The Taylor expansion of $\mathcal{L}$ about $\zeta
=0 $ is obtained to be%
\begin{equation}
\mathcal{L}=-\mathcal{F}-\frac{2}{3}\zeta \left( -2\mathcal{F}\right) ^{3/4}+%
\frac{1}{2}\zeta ^{2}\left( -2\mathcal{F}\right) ^{2/4}-\frac{1}{4}\zeta
^{3}\left( -2\mathcal{F}\right) ^{1/4}+\mathcal{O}\left( \zeta ^{4}\right)
\label{18}
\end{equation}%
which reveals that in the limit $\zeta \rightarrow 0$ the Lagrangian (\ref%
{17}) reduces to Maxwell's linear theory. Finally, the electric field and
electric potential of the theory are given by 
\begin{equation}
E\left( r\right) =\frac{q}{r^{2}}+\zeta \frac{\sqrt{q}}{r}  \label{19}
\end{equation}%
and%
\begin{equation}
U\left( r\right) =\frac{q}{r}-\zeta \sqrt{q}\ln \left( \frac{r}{r_{0}}%
\right) ,  \label{20}
\end{equation}%
respectively. Let us add that, with a magnetic monopole where $\mathcal{F}>0$
one has to consider $\zeta ^{2}$ to be pure imaginary such that $\zeta
^{2}=i\xi ^{2},$ and accordingly the Lagrangian remains real given by%
\begin{equation}
\mathcal{L}=-\frac{16\left( 3\sqrt{2\mathcal{F}}+\xi \left( \xi +\sqrt{\xi
^{2}+4\sqrt{2\mathcal{F}}}\right) \right) \sqrt{2\mathcal{F}}}{3\left( \xi +%
\sqrt{\xi ^{2}+4\sqrt{2\mathcal{F}}}\right) ^{4}}\mathcal{F}  \label{M}
\end{equation}%
in which $\xi $ is a real positive constant.

\section{Nonlinear electrodynamics: Logarithmic interaction}

The simpler case than the previous section is given when we assume the
interaction between the quark and antiquark to be Logarithmic as proposed in 
\cite{2}. Following (\ref{2}), hence, we set the electric potential of the
possible NED model to be ($q>0$) 
\begin{equation}
U\left( r\right) =f\ln \left( \frac{r}{r_{0}}\right)  \label{21}
\end{equation}%
in which $f$ and $r_{0}$ are constant parameters. The corresponding electric
field is obtained to be%
\begin{equation}
E\left( r\right) =\frac{f}{r}  \label{22}
\end{equation}%
and therefore%
\begin{equation}
\mathcal{F}=-\frac{f^{2}}{2r^{2}}.  \label{23}
\end{equation}%
Considering the nonlinear Maxwell equation (\ref{9}) we get%
\begin{equation}
\frac{d\mathcal{L}}{dr}=-\frac{qf}{r^{4}}  \label{24}
\end{equation}%
where we set $C=-q.$ Integrating (\ref{24}) yields%
\begin{equation}
\mathcal{L}=\frac{qf}{3r^{3}}+\mathcal{L}_{0}  \label{25}
\end{equation}%
in which $\mathcal{L}_{0}$ is an integration constant. Finally eliminating $%
r $ between (\ref{23}) and (\ref{25}) we get%
\begin{equation}
\mathcal{L}=\frac{1}{3\zeta ^{2}}\left( -2\mathcal{F}\right) ^{3/2}
\label{26}
\end{equation}%
in which we introduced $f=\sqrt{q}\zeta .$

The NED model obtained in (\ref{26}) belongs to the class of power-law
Maxwell Lagrangian proposed by Hassaine and Martinez in Ref. \cite{5,6}. Its
applications in the physics of the black holes have been demonstrated
through several works \cite{7,8,9,10,11,12,13,14,15,16,17,18}.

\section{Black hole solution}

The NED model obtained in Eq. (\ref{17}) looks rather complicated, however,
as we shall see in this section it manifests interesting features when it is
coupled minimally with gravity. Looking at the roots of (\ref{17}) we recall
that the second term is a confinement term to the original Coulomb law. We
are going to couple this NED model to Einstein's gravity within the
following action%
\begin{equation}
S=\int d^{4}x\sqrt{-g}\left( \frac{\mathcal{R}}{16\pi G}+\mathcal{L}\left( 
\mathcal{F}\right) \right)   \label{27}
\end{equation}%
in which $\mathcal{R}$ is the Ricci scalar, $G=1,$ and $\mathcal{L}\left( 
\mathcal{F}\right) $ is given in Eq. (\ref{17}). As our aim is to obtain
either a particle or a black hole solution within the coupling described by
the action (\ref{27}), we assume the spacetime to be spherically symmetric
with the line element%
\begin{equation}
ds^{2}=-\psi \left( r\right) dt^{2}+\frac{dr^{2}}{\psi \left( r\right) }%
+r^{2}d\Omega ^{2},  \label{28}
\end{equation}%
where $d\Omega ^{2}=d\theta ^{2}+\sin ^{2}\theta d\varphi ^{2}$ is the line
element on the unit 2-sphere. The electromagnetic field of a point charge
sitting at the origin is by definition a radially symmetric electric field
with the only nonzero component expressed in (\ref{5}). We have already
shown that the Maxwell nonlinear Equation i.e., (\ref{7}) is satisfied
provided $E\left( r\right) $ is given by (\ref{19}). What is left is the
Einstein field equation which reads as%
\begin{equation}
G_{\mu }^{\nu }=8\pi T_{\mu }^{\nu }  \label{29}
\end{equation}%
in which $G_{\mu }^{\nu }$ is Einstein's tensor and $T_{\mu }^{\nu }$ is the
NED energy-momentum tensor given by%
\begin{equation}
T_{\mu }^{\nu }=\frac{1}{4\pi }\left( \mathcal{L}\delta _{\mu }^{\nu }-%
\mathcal{L}_{\mathcal{F}}F_{\mu \lambda }F^{\nu \lambda }\right) ,
\label{30}
\end{equation}%
in which $\mathcal{L}_{\mathcal{F}}=\frac{\partial \mathcal{L}}{\partial 
\mathcal{F}}$. The four components of Einstein's field equations are
consistent and practically radius to only $G_{0}^{0}=8\pi T_{0}^{0}$ which
explicitly after some manipulation become%
\begin{equation}
\frac{r\psi ^{\prime }\left( r\right) +\psi \left( r\right) -1}{r^{2}}=-%
\frac{q^{2}}{r^{4}}-\frac{4\zeta q\sqrt{q}}{3r^{3}}  \label{31}
\end{equation}%
which admits the following metric function%
\begin{equation}
\psi \left( r\right) =1-\frac{2M}{r}+\frac{q^{2}}{r^{2}}-\frac{4q\sqrt{q}%
\zeta }{3r}\ln \left( r\right) .  \label{32}
\end{equation}%
Herein, $M$ is an integration constant related to the mass of the black
hole/particle and $q$ is its electric charge. The spacetime is
asymptotically flat and depending on the values of the NED parameter $\zeta $
and the mass and the charge, it is either a singular black hole or a
particle with naked singularity.

\begin{figure}[tbp]
\includegraphics[scale=0.7]{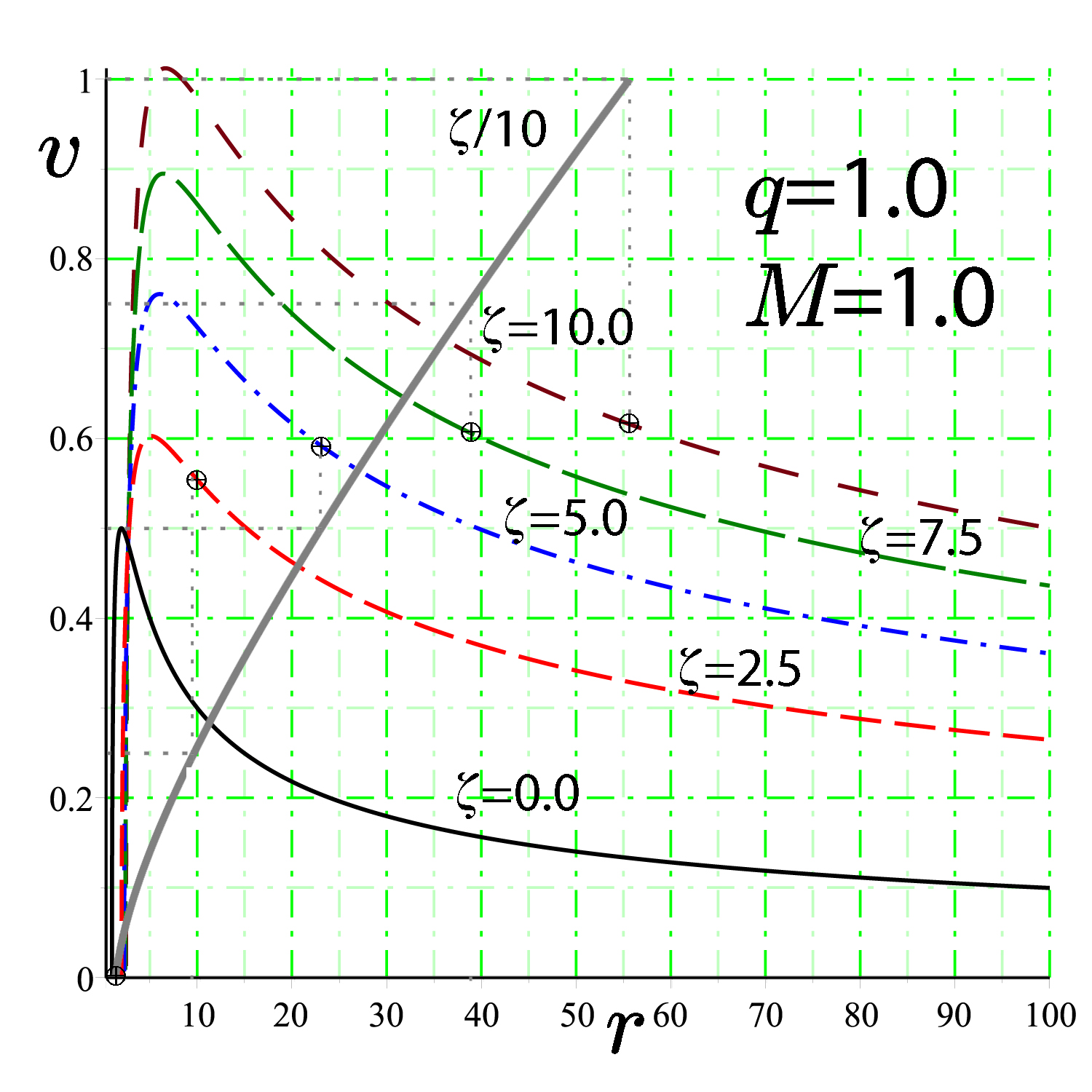}
\caption{Plots of the circular speed $v$ of a star/particle of unit mass
with respect to the distance of the particle/star from the central black
hole $r$ (the center of a galaxy). The value of $\protect\kappa $ is written
on each curve and $q=M=1$. Furthermore, we plotted also $\frac{1}{10}\protect%
\kappa $ against the event horizon for $q=M=1$ (the solid gray curve). Using
this graph we extracted the location of the event horizon on each
speed-radius curve and marked it with the symbol $\oplus .$}
\label{Fig2}
\end{figure}

Next, we assume the spacetime (\ref{28}) with the metric function (\ref{32})
provides the central force for a test particle that orbits the central black
hole at a very large distance in comparison with the horizon of the black
hole. This is in analogy with a distant star in a galaxy moving in a circle
around the central black hole of the galaxy. In the zeroth approximation,
one writes 
\begin{equation}
\psi \left( r\right) =1+2\Phi \left( r\right)  \label{33}
\end{equation}%
in which $\Phi \left( r\right) $ is the Newtonian gravitational potential
for the test particle of unit mass. Therefore, explicitly we get 
\begin{equation}
\Phi \left( r\right) =-\frac{M}{r}+\frac{q^{2}}{2r^{2}}-\frac{2q\sqrt{q}%
\zeta }{3r}\ln \left( r\right) .  \label{34}
\end{equation}%
Here in addition to the traditional terms i.e., $-\frac{M}{r}$ and $\frac{%
q^{2}}{2r^{2}}$ there is an additional term representing the confinement
i.e., $-\frac{2q\sqrt{q}\zeta }{3r}\ln \left( r\right) .$ The latter is
vanishing significantly slower than the other two terms due to the presence
of $\ln \left( r\right) $ in the nominator. Now, the effective gravitational
force is simply given by $F_{c}=-\frac{\partial \Phi \left( r\right) }{%
\partial r}$ which is toward the center of the black hole. We obtain%
\begin{equation}
F_{c}=-\frac{3M-2q^{3/2}\zeta }{3r^{2}}+\frac{q^{2}}{r^{3}}-\frac{%
2q^{3/2}\zeta }{3r^{2}}\ln r  \label{35}
\end{equation}%
which can be equated with centripetal acceleration i.e., $\left\vert
F_{c}\right\vert =\frac{v^{2}}{r}$ in which $v$ is the speed of the test
particle in its orbit. This equation yields an expression for the circular
speed $v,$ given by%
\begin{equation}
v=\sqrt{\frac{3M-2q^{3/2}\zeta }{3r}-\frac{q^{2}}{r^{2}}+\frac{2q^{3/2}\zeta 
}{3r}\ln r}.  \label{36}
\end{equation}%
In Fig. \ref{Fig2}, we plotted the circular velocity $v$ in terms of $r$
(the distance between the particle/star from the center/black hole) for
various values of $\zeta $ and fixed values of $M$ and $q.$ The solid gray
curve is the curve $\frac{1}{10}\zeta $ (vertical axis) versus the radius of
the event horizon (horizontal axis). Using this curve, we have marked the
location of the event horizon for every different curve with symbols $\oplus 
$ on each curve. This figure displays a great improvement in the
speed-radius curve anomaly.

\subsection{Physical properties of the black hole}

The black hole solution (\ref{32}) is considered to be the Reissner-Nordstr%
\"{o}m black hole with a correction term proportional to $\zeta $. The black
hole is asymptotically flat and depending on the parameters namely $\zeta ,$ 
$M$, and $q$ admits: i) two distinct (i.e., inner and outer) horizons for $%
M>M_{c}$, ii) one degenerate horizon for $M=M_{c}$ and iii) no horizon for $%
M<M_{c}$, in which%
\begin{equation}
M_{c}=\frac{q^{2}}{\Delta }+\frac{2\zeta q^{3/2}}{3}\left( 1-\ln \Delta
\right)  \label{R1}
\end{equation}%
and 
\begin{equation}
\Delta =\frac{2\zeta q^{3/2}}{3}\left( 1+\sqrt{1+\frac{9}{4q\zeta ^{2}}}%
\right) ,  \label{R2}
\end{equation}%
provided $M_{c},M\geq 0.$ While the general expression of the inner and
outer horizon is not analytically available, in the extremal black hole
configuration where $M=M_{c}>0$, one obtains%
\begin{equation}
r_{+}=r_{-}=\Delta .  \label{R3}
\end{equation}%
Furthermore, the solution is singular at the center of the black hole and
its Ricci scalar is simply given by%
\begin{equation}
\mathcal{R}=\frac{4q^{3/2}}{3r^{3}}\zeta .  \label{R4}
\end{equation}%
Next, we check the energy conditions namely the null energy condition (NEC),
the weak energy condition (WEC), and the strong energy condition (SEC) (see
for instance \cite{R2}). Considering the energy-momentum tensor of the
matter field supporting the black hole given by%
\begin{equation}
T_{\mu }^{\nu }=diag\left[ -\rho ,p_{r},p_{\theta },p_{\varphi }\right]
\label{R5}
\end{equation}%
in which 
\begin{equation}
\rho =-p_{r}=p_{\theta }=p_{\varphi }=\frac{1}{8\pi }\left( \frac{q^{2}}{%
r^{4}}+\zeta \frac{4q^{3/2}}{3r^{3}}\right) ,  \label{R6}
\end{equation}%
the NEC ($\rho +p_{i}\geq 0$), the WEC ($\rho +p_{i}\geq 0$ and $\rho \geq 0$%
) and the SEC ($\rho +\sum p_{i}\geq 0$) are all satisfied provided we
impose $\zeta \geq 0.$

Another interesting aspect of black holes is their thermal stability. To
check the thermal stability of the black hole solution (\ref{32}), we first
impose $M\geq M_{c}\geq 0$ and calculate the Hawking temperature at its
event horizon $r=r_{+}$. Hence, the Hawking temperature is given by%
\begin{equation}
T_{H}=\frac{1}{4\pi }\psi ^{\prime }\left( r_{+}\right) =\frac{1}{4\pi r_{+}}%
-\frac{q^{2}}{4\pi r_{+}^{3}}-\zeta \frac{q^{3/2}}{3\pi r_{+}^{2}}.
\label{R7}
\end{equation}%
Furthermore, the black hole entropy is given by ($G=1$) 
\begin{equation}
S=\frac{1}{4}A\left( r_{+}\right) =\pi r_{+}^{2},  \label{R8}
\end{equation}%
where $A\left( r_{+}\right) $ is the area of the black hole at the event
horizon. Using (\ref{R7}) and (\ref{R8}), the heat capacity of the black
hole for the constant charge is calculated to be%
\begin{equation}
C_{q}=\left( T_{H}\frac{\partial S}{\partial T_{H}}\right) _{q}=\frac{2\pi
r_{+}^{2}\left( 3r_{+}^{2}-4q^{3/2}\zeta r_{+}-3q^{2}\right) }{%
-3r_{+}^{2}+8q^{3/2}\zeta r_{+}+9q^{2}}.  \label{R9}
\end{equation}%
The latter expression for the heat capacity together with the Hawking
temperature (\ref{R7}) reveal that the interval for $r_{+}$ where both are
positive is given by%
\begin{equation}
r_{1}<r_{+}<r_{2}  \label{R10}
\end{equation}%
where $r_{1}$ and $r_{2}$ are so-called the first and the second transition
points, given by 
\begin{equation}
r_{1}=\frac{2\zeta q^{3/2}}{3}\left( 1+\sqrt{1+\frac{9}{4q\zeta ^{2}}}%
\right) =\Delta  \label{R11}
\end{equation}%
and%
\begin{equation}
r_{2}=\frac{4\zeta q^{3/2}}{3}\left( 1+\sqrt{1+\frac{27}{16q\zeta ^{2}}}%
\right) .  \label{R12}
\end{equation}%
At $r_{+}=r_{1}$ both $T_{H}$ and $C_{q}$ are zero that represents the
extremal black hole, however, at $r_{+}=r_{2}$ the Hawking temperature
becomes maximum and the heat capacity diverges. The latter is the Davies
point \cite{Da1,Da2}.

\subsection{Smarr formula and the first law of thermodynamics}

For the Schwarzschild black hole the mass $M,$ the Hawking temperature $%
T_{H},$ and the entropy $S$ satisfy a relation given by $M=2TS$ which is
known as the Smarr relation/formula. In general Smarr formula is a
connection between gravity and the thermodynamics of the black hole. For the
Reissner-Nordstr\"{o}m black hole in Einstein-Maxwell theory, the Smarr
formula is simply the generalization of Schwarzschild's one given by $%
M=2TS+Q\Phi ,$ in which $Q$ is the electric charge and $\Phi $ is the
electric potential. Unlike the linear Maxwell theory, in the nonlinear
electrodynamics for the black hole solution,\ a Smarr formula is not so
straightforward. This is due to the nonzero trace of the energy-momentum
tensor. For the black hole solution in a nonlinear electrodynamics model
that asymptotically behaves as the Reissner-Nordstr\"{o}m black hole, one
may use the definition of the Komar mass/energy to construct the
corresponding Smarr formula \cite{KS}. Moreover, a geometric derivation of
the generalized Smarr formula for the black holes in nonlinear
electrodynamics has also been introduced in \cite{GS}. In both approaches
the condition of having the nonlinear term in the metric function going to
zero faster than $\frac{1}{r}$ in the limit $r\rightarrow \infty $ is
crucial. As we have already mentioned the black hole solution (\ref{32}) is
asymptotically flat, however, the NED term is not going to zero faster than
the mass term. This makes the solution ambiguous for the definition of the
mass and other thermodynamic quantities in order to establish the Smarr
formula in consistency with the first law of thermodynamics. To resolve this
problem, we follow the method of renormalization introduced in \cite{R3}
(see also \cite{R4,R5}). Following \cite{R3}, we assume that the entire
spacetime is enclosed in a sphere of radius $r_{0\text{ }}$and upon
introducing a new mass constant we rewrite the metric function in the
following form%
\begin{equation}
\psi \left( r\right) =1-\frac{2\mathcal{M}}{r}+\frac{q^{2}}{r^{2}}-\frac{4q%
\sqrt{q}\zeta }{3r}\ln \left( \frac{r}{r_{0}}\right)  \label{R14}
\end{equation}%
where the relation between $M$ and $\mathcal{M}$ is given by 
\begin{equation}
\mathcal{M}=\frac{2q\sqrt{q}\zeta }{3}\ln r_{0}+M.  \label{R15}
\end{equation}%
Unlike $M$ which represents the gravitational mass, $\mathcal{M}$ stands for
the total electromagnetic and gravitational energy enclosed inside the
sphere of radius $r_{0}.$ In the limit of $r\rightarrow \infty ,$ one
assumes $r_{0}\rightarrow \infty $ such that the ration $\frac{r}{r_{0}}=1$
which results in the nonlinear term vanishes. At the event horizon 
\begin{equation}
\mathcal{M}\left( r_{+},q,\zeta ,r_{0}\right) =\frac{r_{+}}{2}+\frac{q^{2}}{%
2r_{+}}-\frac{2q^{3/2}}{3}\zeta \ln \frac{r_{+}}{r_{0}},  \label{R16}
\end{equation}%
which in terms of the entropy $S$ can be rewritten as%
\begin{equation}
\mathcal{M}\left( r_{+},q,\zeta ,r_{0}\right) =\frac{\sqrt{S}}{2\sqrt{\pi }}+%
\frac{q^{2}\sqrt{\pi }}{2\sqrt{S}}-\frac{2q^{3/2}}{3}\zeta \ln \left( \frac{%
\sqrt{S}}{r_{0}\sqrt{\pi }}\right) .  \label{R17}
\end{equation}%
From the electric field ansatz (\ref{5}) and the Bianchi identity, one finds 
$\mathbf{F}=d\mathbf{A}$ where $\mathbf{A}=A\left( r\right) dt$ is the
electric gauge potential and%
\begin{equation}
A\left( r\right) =\frac{q}{r}-\zeta \sqrt{q}\ln \left( \frac{r}{r_{0}}\right)
\label{R18}
\end{equation}%
where the constant potential is considered to be $\zeta \sqrt{q}\ln \left(
r_{0}\right) $. Hence the electric potential of the black hole is defined to
be 
\begin{equation}
\Psi =A\left( r_{+}\right) =\frac{q}{r_{+}}-\zeta \sqrt{q}\ln \left( \frac{%
r_{+}}{r_{0}}\right) .  \label{R19}
\end{equation}%
Furthermore, differentiating (\ref{R17}) implies 
\begin{equation}
d\mathcal{M}=\left( \frac{\partial \mathcal{M}}{\partial S}\right) _{q,\zeta
,r_{0}}dS+\left( \frac{\partial \mathcal{M}}{\partial q}\right) _{S,\zeta
,r_{0}}dq+\left( \frac{\partial \mathcal{M}}{\partial \zeta }\right)
_{S,q,r_{0}}d\zeta +\left( \frac{\partial \mathcal{M}}{\partial r_{0}}%
\right) _{S,q,\zeta }dr_{0}  \label{R20}
\end{equation}%
yielding the first law of thermodynamics%
\begin{equation}
d\mathcal{M}=T_{H}dS+\Psi dq+K_{\zeta }d\zeta +K_{r_{0}}dr_{0}.  \label{R21}
\end{equation}%
From (\ref{R7}) and (\ref{R19}), it can directly be confirmed that%
\begin{equation}
T_{H}=\left( \frac{\partial \mathcal{M}}{\partial S}\right) _{q,\zeta
,r_{0}}=\frac{1}{4\sqrt{\pi S}}-\frac{\zeta q^{3/2}}{3S}-\frac{\sqrt{\pi }%
q^{2}}{4S^{3/2}},  \label{R22}
\end{equation}%
and%
\begin{equation}
\Psi =\left( \frac{\partial \mathcal{M}}{\partial q}\right) _{S,\zeta
,r_{0}}=\frac{q\sqrt{\pi }}{\sqrt{S}}-\zeta \sqrt{q}\ln \left( \frac{\sqrt{S}%
}{r_{0}\sqrt{\pi }}\right)  \label{R23}
\end{equation}%
are the Hawking temperature and the electric potential. On the other hand%
\begin{equation}
K_{\zeta }=\left( \frac{\partial M}{\partial \zeta }\right) _{S,q}=-\frac{2}{%
3}q^{3/2}\ln \left( \frac{\sqrt{S}}{r_{0}\sqrt{\pi }}\right)  \label{R24}
\end{equation}%
and%
\begin{equation}
K_{r_{0}}=\left( \frac{\partial \mathcal{M}}{\partial r_{0}}\right)
_{S,q,\zeta }=\frac{2q^{3/2}\zeta }{3r_{0}}  \label{R25}
\end{equation}%
are the conjugate quantities of the nonlinear parameter $\zeta $ and the
renormalization parameter $r_{0}$. Finally, the modified Smarr formula is
obtained to be%
\begin{equation}
\mathcal{M}=2TS+q\Psi +K_{r_{0}}r_{0}-\frac{1}{2}K_{\zeta }\zeta .
\label{R26}
\end{equation}%
There are other works on the Smarr formula which we refer to \cite{R6}\ and
the references therein.

\section{Conclusion}

We introduced two NED given in Eqs. (\ref{17}) and (\ref{26}) whose electric
potential, due to a point electric charge, is in the form of a confinement
potential between a quark and an antiquark. The second NED model obtained in
Eq. (\ref{26}) is based on the potential (\ref{2}) proposed in \cite{2}. It
is known to be in a class of NED called the power-law Maxwell NED model \cite%
{5,6}. On the other hand, the first NED model introduced in (\cite{17}) is
based on the confinement potential expressed in (\ref{3}) which looks to be
a combination of the Cornell potential and the Logarithmic one. Unlike the
other model, this NED model has not been reported before and is new in this
sense. Its exact form given in Eq. (\ref{17}) looks to be complicated in
comparison with the other models, however, its actual behavior is rather
simpler than most of the models existing in the literature. This can be
easily seen from the electric field solution of Maxwell's equation which is
simply the Coulomb field plus a term proportional to ($\frac{1}{r}$) as
given in Eq. (\ref{19}). Moreover, we investigated the effect of the
nonlinear term in the spacetime solution of the gravity coupled to this NED.
The spacetime that emerged from such nonlinear coupling is either a singular
black hole or a naked singular particle. In the case of a black hole, we
examined the circular velocity of a distant test particle orbiting the black
hole, in analogy with a star orbiting the central black hole in a large
galaxy. Being aware of the speed-radius anomaly (see \cite{G1,G2,G3,G4,G5}
and the references therein) of such distant stars, we have shown that in the
zeroth approximation, the Newtonian potential of the black hole acting on
the distant star causes a speed-radius curve which is significantly improved
in comparison with the Reissner-Nordstr\"{o}m central black hole. Since the
rotational effects of the central black hole have not been counted we don't
expect an exact match with the empirical results but at least it shows a
weaker confinement when it is compared with the case without correction.
Furthermore, we have studied the physical properties of the black hole (\ref%
{32}).

\end{document}